# Tuneable and biodegradable poly(ester amide)s for disposable facemasks

Esteban Alvarez Seoane[1]*, Alessandro Cattaneo[2]*, Fabien Neuenschwander[2], Lucien Blanchard[2], Tatiana Nogueira Matos[1], Laure Jeandupeux[1], Gianni Fiorucci[1], Maryam Tizgadam[2], Kelly Tran[2] Pierre-Louis Sciboz[2], Luce Albergati[2], Jérôme Charmet[1‡], Roger Marti[2‡], Stefan Hengsberger[2‡]

1. Haute Ecole Arc, HES-SO University of Applied Sciences and Arts of Western Switzerland, 2002 Neuchâtel, Switzerland.
2. Haute Ecole d'Ingénierie et d'Architecture de Fribourg, HES-SO University of Applied Sciences and Arts of Western Switzerland, 1700 Fribourg, Switzerland

   * Joint first authors

   ‡ Corresponding authors (jerome.charmet@he-arc.ch, stefan.hengsberger@hefr.ch, roger.marti@hefr.ch)

## Abstract

The widespread use of disposable facemasks during the COVID-19 pandemic has led to environmental widespread concern due to microplastic pollution. Biodegradable disposable facemasks are a first step to reducing the environmental impact of pandemics. In this paper we present high-performance facemask components based on novel poly(ester amide) (PEA) grades synthesized from biosourced materials and processed into non-woven facemask components. PEA based polymers present an excellent compromise between mechanical performance and biodegradability. Importantly, the properties of the PEA can easily be tuned by changing the ratio of the ester and amides, or variation of diol and diacid part. We synthesized seven polymers which we optimized for biodegradability and processability. Among them, two grades combined electrospinning process compatibility with full degradation within 35 days, using a normalized biodegradation test. The ultra-thin filters thus developed were evaluated for performance on a custom-made characterization bench. The filters achieved a microparticle capture efficiency and breathability comparable to commercial filters. Another PEA grade was optimized to reach optimal viscothermal properties that made it compatible with solvent-free melt-spinning process as demonstrated with continuous fibres production. Overall, our environmentally friendly solution paves the way for the fabrication of high-performance fibres with excellent biodegradability for the next generation facemasks.

## Introduction

During the COVID-19 pandemic, facemasks proved to be an effective way to limit the spread of the virus[1,2]. Commercial medical facemasks are high-tech products which must meet many requirements in terms of hygiene, efficiency, and cost. In general, they are multi-layered and include a high-performance filter layer. A recently published communication highlights the environmental impact resulting from disposable facemasks[3]. According to the authors, the microplastics originating from the standard masks, made of polypropylene and polyethylene, could significantly aggravate the global plastic pollution. Other studies also confirm the potential environmental damage that results from improper disposal of facemasks[4]. This ecological issue raises new challenges to the textile industry.

Electrospinning[5,6] is a highly promising technique that was shown to not only improve facemask performance, but also reduce the amount of polymer used compared to conventional mask fabrication processes[7–9]. However, even though this fabrication method is more environmentally friendly, the solution is not ideal as it still produces plastic waste. Instead, one of the most promising solutions is to fabricate the facemasks using biodegradable polymers.

This class of polymers has received considerable attention for biomedical applications[10,11], including for facemasks. Instances of electrospun biodegradable facemasks made of Polylactic acid (PLA)[12,13]; cellulose[14,15] chitosan[16,17] and other materials[18,19] were reported recently. Among them, PLA[12] and poly(butylene succinate)[20] electrospun air filters were shown to exhibit high filtration efficiency and good biodegradability.

Among the biodegradable polymers available, poly(ester amide)s (PEA) appear as a promising candidate as they harbour the high thermal stability, high elastic modulus and high tensile strength of polyamides combined with the good degradability of polyesters[21]. Another interesting feature of PEAs is their properties tuneability. This explains why poly(ester amide)s grades have received a lot of attention for biomedical applications[22,23]. Even though the fabrication of high-grade filters, made by electrospinning of PEA fibres was demonstrated[21,24] there is no report, to the best of our knowledge, of PEA based facemasks that demonstrate excellent biodegradability combined with filtration and breathability on par with the performance of commercial facemasks.

In this paper we present and fully characterise novel biodegradable facemasks components made of PEA fibres. In brief, we synthesised seven biosourced poly(ester amide) grades and evaluated them for processing into non-woven fibres. By varying the ratio of ester and amides and through the tuning of diol and diacid, we systematically optimised PEA grades for biodegradability and two non-woven fibres fabrication processes (Figure 1). Selected candidates underwent a normalised biodegradation test, and two polymers were fully degraded in less than 35 days, including one that degraded within 20 days, which is comparable to cellulose. To evaluate the performance of the biodegradable electrospun filters, we realised a custom bench to measure filtration efficacy and breathability. Compared to commercial filters, our ultra-thin filters demonstrate similar filtration efficiency and breathability. Finally, we demonstrate that one of our PEA grades is compatible with solvent-free melt spinning process for the fabrication of outer layers fabric. In particular we optimised the fabrication process to enable continuous fibre formation on a custom-made rig. Overall, our results pave the way for the development of high-performance biodegradable facemasks based on biosourced PEA.

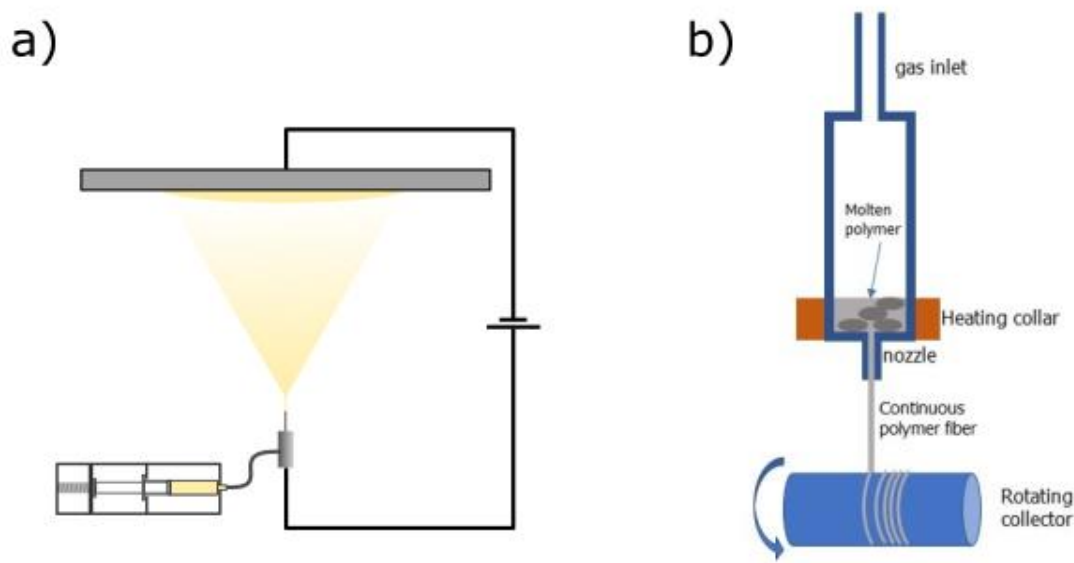

**Figure 1**. *The poly(ester amide)s synthesised and presented in the manuscript were fine tuned to enable their processing by a) electrospinning and b) melt-spinning to allow for the fabrication of high performance mask filters and fibers compatible with mask outer layers respectively.*

# Results

## Polymer Synthesis & Characterization

As shown in Figure 2a, a series of poly(ester amide)s **1-6** were synthesized by polycondensation from biobased raw materials such as diols (1,4-butanediol, 1,6-hexanediol and 1,10-decanediol), diesters (dimethyl adipate and dimethyl 2,5-furandicarboxylate DMFD) and a bisamide-siol building block prepared from 1,4-butanediamine and caprolactone[25–29]. The ratio of unique bisamide-diol building block was varied to prepare polymers with different ester-amide content. GPC measurements were performed on the polymers to measure the MW, Mn and PDI as reported in Table 1.

**Table 1**: *Overview of Biobased Poly(ester amide)s by Polycondensation*

| Polymer | Diol | Diester | Ester amide ratio | MW [g/mol] | Mn [g/mol] | PDI |
|---|---|---|---|---|---|---|
| 1 | 1,4-butanediol | *Dimethyl adipate* | 50/50 | 17'700 | 4'800 | 3.5 |
| 2 | 1,4-butanediol | *DMFD* | 50/50 | 11'171 | 5'558 | 2.01 |
| 3 | 1,10-decanediol | *DMFD* | 50/50 | 36'222 | 16'078 | 2.25 |
| 4 | 1,6-hexanediol | *DMFD* | 75/25 | 51'836 | 20'346 | 2.487 |
|   | 1,6-hexanediol | *DMFD* | 75/25 | 48'957 | 20'888 | 2.344 |
| 5 | 1,6-hexanediol | *DMFD* | 50/50 | 7'512 | 3'699 | 2.031 |
| 6 | 1,6-hexanediol | *DMFD* | 25/75 | 37'528 | 17'220 | 2.179 |

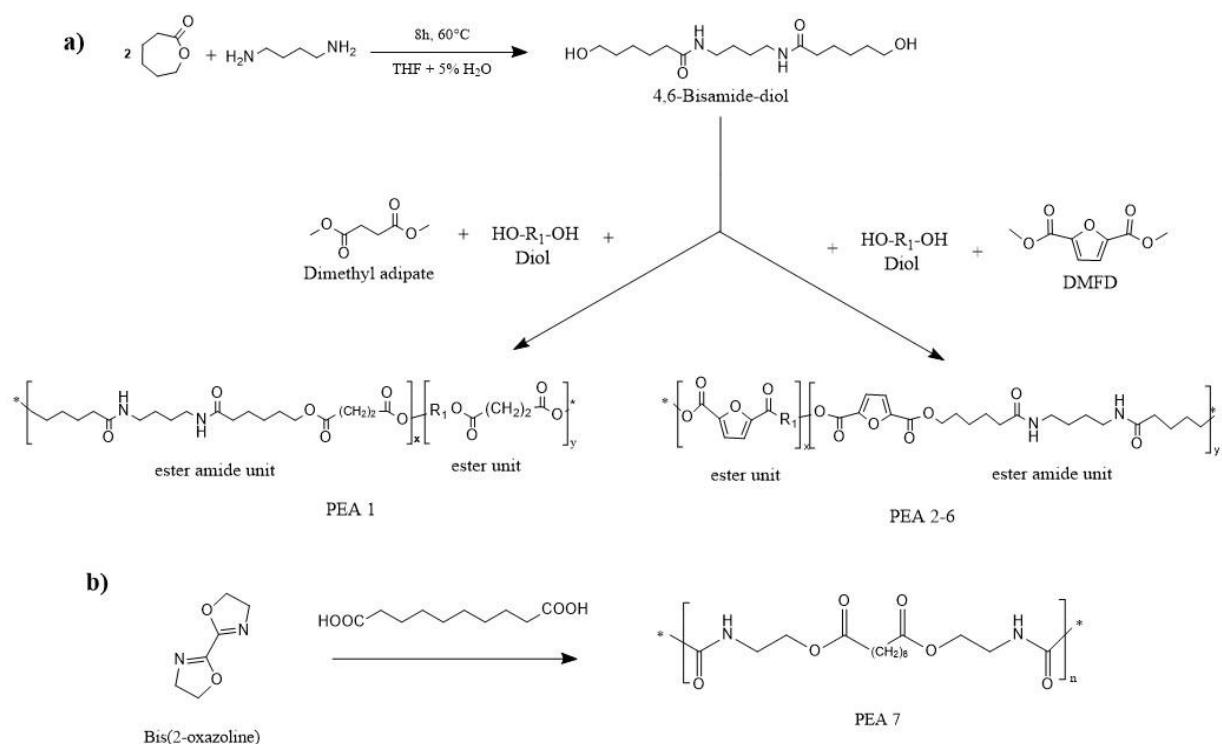

**Figure 2.** Synthesis of poly(ester amide)s grades. a) Biobased poly(ester amide)s by polycondensation with dimethyl adipate (PEA **1**), and with Dimethyl 2,5-furandicarboxylate (PEA **2-6**). b) Bis(oxazoline)-Based poly(ester amide) **7** by polyaddition.

In addition, poly(ester amide) **7** based on bis(oxazoline) and sebacic acid was prepared by polyaddition reaction (Figure 2b)[30,31]. The first synthesis (polymers **1,2**) were performed with 1,4-butanediol at an ester-amide ratio of 50%:50%. This initial selection was done due to the bio-based origin of this diol and the lower boiling point compared to the other candidates (1,6-hexanediol and 1,10-decanediol). The tests performed showed that low molecular weight polymers were synthesized due to sublimation of the oligomers formed during the polycondensation reaction and thus leading to low molecular weight polymers **1** and **2**. Polymer **3** synthesized with 1,10-decanediol was straightforward and showed a higher molecular weight in comparison to the first polymers synthesised. However, the cost of this diol encouraged us to select 1,6-hexanediol for further trials and the preparation of polymers **4**, **5**, and **6**.

The summary of the thermal characterization of polymers **1-7** is shown in Table 2. The thermograms of the different polymers shows a decrease of the glass transition temperature with an increase of the diol aliphatic chain length (Table 2 and Figure S1 in SI). This phenomenon is linked to the flexibility of the polymers. The longer the aliphatic chain of the diol the easier it is for the polymer to pass from a glassy state to an amorphous state due to the increased movement possibilities offered by the polymer's chains. A cold crystallization peak can be seen in all the thermograms before melting. The introduction of hydrogen-bond prone amide segment allows this thermal event in the polymers. As for the Tg, the flexibility of the polymer increases with the length of aliphatic chain of the diol, thus reducing the energy needed to pass from an amorphous state to a crystal[25,32].

*Table 2: Overview of Thermal Data (DSC) for the seven polymer grades.*

| Polymer | Tg [°C] | Cold cryst [°C] | Mp 1 [°C] | Mp 2 [°C] |
|---|---|---|---|---|
| 1 | - 38.4 | - | 73.71 | 133.3 |
| 2 | 27.5 | 111.6 | 134.4 | - |
| 3 | 12.7 | 71.9 | 139.7 | - |
| 4 | 18.1 | - | 119.8 | - |
| 5 | 26.0 | 105.5 | 137.5 | - |
| 6 | 27.5 | - | 157.8 | 171.4 |
| 7 | 17.8 | 152.0 | 172.75 | - |

The thermal analysis was also performed on polymers synthesized from 1,6-hexanediol with different ester-amide ratio (Table 2 and Figure S2 in SI). From the thermograms one observes that Tg increases with the amide content from an ester amide ratio of 75-25 (polymer **4**) to the 50-50 (polymer **5**). After this increase the Tg stabilises at around 27°C. Another observation is that a cold crystallization is present at the 50-50 ester amide ratio but not in the other two polymers. Finally, the presence of two distinct melting peaks in the PEA with an ester amide ratio of 25-75 (polymer **6**) indicating the melting of the ester and the amide segments. In the other polymers the ester melting was not observed.

## Solubility tests and initial electrospinning trials

The polymer solubility was tested in different solvents (Table S1 in SI). For the first tests of solubility the polymers **3**, **4** and **6** synthesized from 1,4-butanediol, 1.6-hexanediol and 1,10-decanediol with a 50% of hard segment were chosen. In most cases the polymers present solubility in alcohols and halogenated solvents and are not soluble in carbonates or N-methylmorpholine N-Oxyde. The results also show that there is no specific solubility pattern. Where mixture of solvents was used, the results presented even greater variability. Methanol represents an exception as most tests were inconclusive when it was present. The solubility of the polymers is highly dependent on the molecular weight, polar forces, hydrogen bonding and dispersion forces[33]. Thus, the solubility of polymer has to be evaluated and fine-tuned for each batch synthesized.

From these initial observations, a selection of possible solvents for electrospinning was performed. The initial screening was performed on the polymers synthesized from 1,6-hexanediol and 1,10-decanediol. Polymers produced from 1,4-butanediol were not tested due to low molecular weights. The tests performed are summarized in the table 3.

The results of this screening tests suggest that HFIP is the best candidate amongst the solvents tested. High boiling point solvents (dimethylcarbonate (DMC), Phenyl-ethanol, benzyl alcohol) were selected due to their non-hazardous nature and for the fact that they are often found in nature (fruits) making them potentially bio-based. The problem with these solvents is their low vapor pressure which lower their evaporation rate compared to solvents such as hexafluoroisopropanol (HFIP) and dichloromethane (DCM). The rapid removal of solvent is important to obtain good electrospinning results. We also performed solubility tests in binary systems that combine a high boiling point with a low boiling point solvent (Methanol, Ethanol, DCM and Chloroform). Our idea behind this was to maintain the good solubility provided by the high boiling point solvents, while improving evaporation during electrospinning due to the presence of low boiling point solvent. The results of these tests were mostly electrospraying or gelatinous mixture of solvent and polymer on the collector. Overall, the tests performed with HFIP showed the best results and enabled the deposition of fibrous material. This solvent was thus selected for further processing.

*Table 3:* Summary of initial electrospinning tests and conditions

| Polymer | Solvent | Conc. [%wt.] | Voltage [kV] | Distance [cm] | Flow rate [µL / min] | Collector speed [rpm] | Result |
|---|---|---|---|---|---|---|---|
| 3 | HFIP | 12 | 17 | 12 | 21 | 50 | Fibers |
| 4 | HFIP | 8 | 17 | 12 | 31 | 50 | Fibers |
| 4 | Chloroform / phenyl ethanol | 10 | 14 | 15 | 5 | 100 | Not continuous filament |
| 4 | Chloroform / benzyl alcohol | 15 | 9 | 15 | 5 | 100 | Spraying |
| 4 | Chloroform / phenyl ethanol | 12.5 | 11 | 15 | 35 | 100 | Not continuous filament |
| 4 | DCM / benzyl alcohol | 13 | 9 | 15 | 35 | 100 | Not continuous filament |
| 4 | Chloroform /DCM | 6.7 | 25 | 15 | 20 | 100 | Spraying |
| 6 | HFIP | 10 | 17 | 13 | 21 | 50 | Fibers |
| 6 | Chloroform / methanol | 6.5 | 8 | 5 | 31 | 100 | Spraying |
| 6 | Chloroform / ethanol | 6.7 | 25 | 15 | 20 | 100 | Spraying |
| 6 | DCM / ethanol | 6.7 | 25 | 11 | 35 | 100 | Spraying |
| 6 | DCM / benzyl alcohol | 6.7 | 25 | 11 | 20 | 180 | Not continuous filament |
| 6 | Chloroform / DMC | 6.7 | 25 | 11 | 20 | 180 | spraying |
| 6 | DCM / Methanol | 6.7 | 25 | 11 | 20 | 180 | Spraying |
| 6 | DCM / phenyl ethanol | 6.7 | 10 | 15 | 30 | 180 | Not continuous filament |

## Biodegradation tests

Figure 3 shows the biodegradation results for three PEA grades based on the norm ISO 14855-1[34]. The method reported in the norm involves the measurement of carbon dioxide as a function of time allowing to determine the degradation of the materials in comparison to cellulose reference. A target value of 90% is considered as a total decomposition. Tests showed a rapid degradation of the two PEA grades polymer **1** and polymer **7**. These polymers were completely degraded after less than 35 days, with polymer **1** following the degradation curve of cellulose and a full degradation after about 20 days. The polymer **4** grade presents a slower rate of degradation with a plateau after 45 days followed by an increase between 75 days and 105 days. After 105 days the polymer slowly continues the degradation until a value of 70% after 180 days. The difference in the degradation is attributed to the different chemical structure and molecular weight of the polymers.

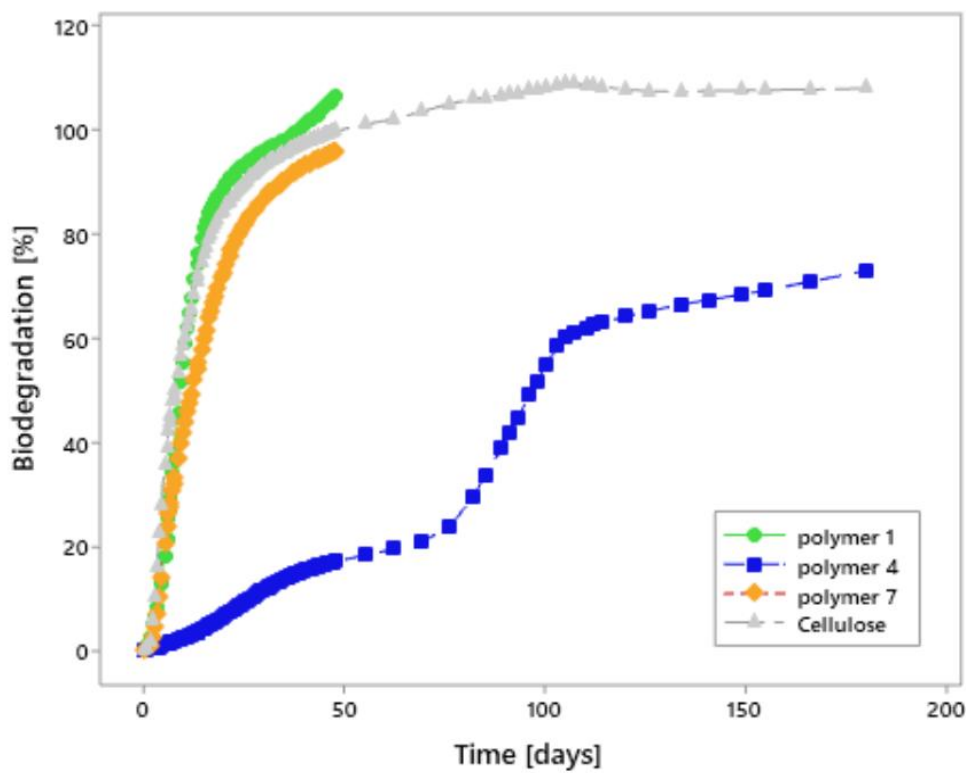

*Figure 3. Biodegradability tests based on the norm ISO 14855-1 of three selected poly(ester amide) grades presented in this paper. The degradation is tested with cellulose as a reference material. A target value of 90% is considered as a total degradation. Polymer **1** achieved a degradation on par with cellulose, with a degradation within 20 days. Polymer **7** was fully degraded after 35 days. Biodegradation was observed for polymer **4**, albeit it was slower than for the other two polymers.*

### Electrospinning and parameter optimisation

Based on the above results, polymer **7**, which performed well in the biodegradation test (Figure 3) and produced fibrous materials when dissolved in HFIP during the electrospinning screening tests, was selected to create high-performance filters. Different polymer concentrations were tested for fibres quality by systematically varying the flow rate, the collector distance and voltage (see Design of Experiment section and Table S2 in SI). Figures 4a shows the heatmaps that represent the quality of the electrospun fibers obtained from 10 wt.% solution of polymer **7** in HFIP. The green areas show high-quality, homogeneous fibers, while the red areas show very poor-quality fibres or electro-spraying. Figure 4a shows that slower flow rate and higher voltage/distance ratio improves the quality of the fibers. In contrast, the heatmap for 12.5 wt.% solution of polymer **7** in HFIP is (Figure S3 in SI) is almost entirely green under the same conditions.

The stark contrast between the results for the 10 wt.% and the 12.5 wt.% highlights the fact that the most important parameter for successful fibre formation is the polymer concentration in the solvent. Indeed, this parameter influences both viscosity and the required time for fibre strands to dry out into solid polymer. This observation is in line with other studies[35]. Then, increasing the voltage and reducing the collector distance also shows a clear improvement of fibre quality, although one should be wary of the influence on process speed and deposition area. For the feed-rate, a balance should be found between increased process speed (high feed-rate) and better-quality fibres (low feed-rates).

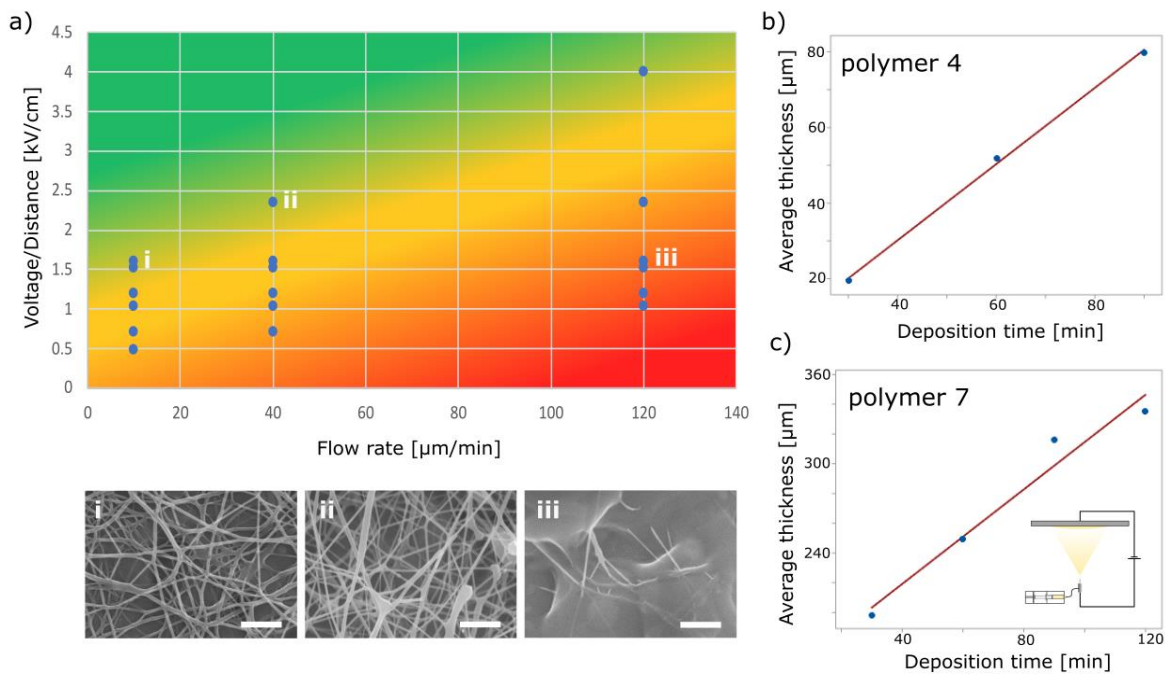

*Figure 4.* Electrospinning optimisation and characterisation. a) Fibre quality heatmap as function of the flowrate and the ratio of voltage to distance to collector. Data for 10wt.% solution of polymer **7** in HFIP. Insets show scanning electron micrographs of fibers obtained after deposition under conditions i, ii and iii. Filter thickness as function of the deposition time for polymer **4** (b) and polymer **7** (c). A clear linear correlation is observed between deposition time and polymer layer thickness for each polymer with $R^2$ value of 99.8% and 96.2% for polymer **4** and **7** respectively.

## Filter fabrication and characterisation

### Correlation between deposition time and layer thickness

The capture efficiency and the breathability of filter depends on fibre size and density, and on the overall filter thickness. The first two parameters were evaluated using scanning electron microscopy and for the latter, we used a confocal microscope.

Layer thickness is dependent on feed rate, collector- emitter distance and importantly deposition time. Using optimized electrospinning parameters described above, we prepared samples and analysed samples as described in the materials section. It should be noted that when using a static planar collector, the thickness is location dependent, with fibres depositing faster at the centre of the pattern than at the edges[36]. Therefore, comparison between samples were made on sections cut at the same location each time as explained in the Methods section.

Even though the residual charges buildup on the collected fibres tends to repel the similarly charged jet which limits the maximum thickness of the layer, our data (Figure 4b and 4c, for polymer **4** and **7** respectively) show that we are still in the linear regime despite deposition times over what is needed for the fabrication of our filters, as shown below. This ensures that we simply control filter thickness by varying the deposition time.

### Filter performances measurement

**Figure 5d** shows a sketch of the characterization bench developed to measure the filtering efficacy of the filters. An optical image of the bench is available in Figure S4, in SI. A magnetic agitator is used to create an aerosol of Teflon particles that directly passes to the particle counter (channel 1). Once the particle flow intensity is determined, the particle stream is directed to filter (channel 2) and the resulting particles passing through are counted, thus enabling differential measurement.

Since the particle detector allows for an independent analysis of 1 and 3 µm particles, two norms (95% absorption of 3 µm particles OR 70% absorption of 1 µm particles) currently in use in Europe (95%/3µm) and proposed by Swiss hospitals (70%/1µm)[37] could be evaluated.

For the breathability characterization, the aerosol generator is removed, and the particle counter is replaced by a vacuum pump. An air flow controller (PFM750S-F01-F, Distrelec) and a differential pressure detector (Manometer Testo 512, 0-20hPa) are then added to the circuit. For the analysis the pressure drop through the filter was evaluated for an air flow range between 0 and 14 lt/min.

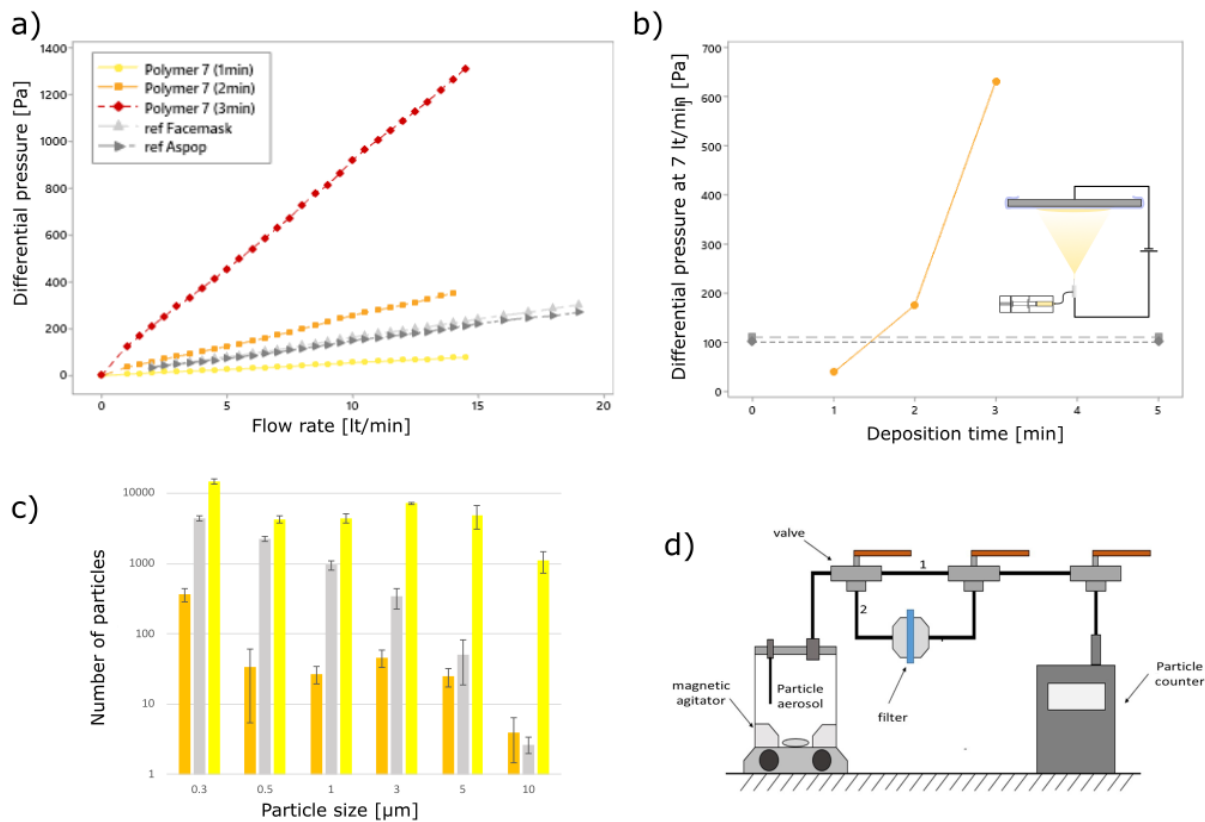

*Figure 5.* Electrospun fibre characterisation. a) Pressure drop for three electrospun filters (polymer 7) with 1,2 and 3 min electro-spinning deposition time with respect to two commercial facemasks for different air flow rates. (filter & outer layer for all tested samples) b) Optimisation of filter electrospinning deposition time for polymer 7. In this figure the pressure drop values are compared for an air flow rate of 7 lt/min and the breathability compared to commercial filters (two horizontal lines). c) Absorption test of polymer 7 filters (electrospun filter on outer mask layer) in comparison to the commercial reference Facemask (filter & outer layer). The number of transmitted microparticles for electro-spinning deposition time between 1 min and 2 min are in general respectively above and below the values obtained using commercial filter. d) Principle of the filter performance test: Teflon microparticles are inserted in a magnetic agitator to generate a particle aerosol. A particle counter with an integrated pump analyses the transmitted microparticles. The particle flow can be alternatively directed through channel 1 for a control of the particle flow intensity and channel 2 to analyse the absorption through the filter. For the measurement of the breathability the particle counter is replaced through a vacuum pump with added flowmeter. The differential pressure is measured on both sides of the filter applying and a pure air flow is applied.

Varying the deposition time to control filter thickness allowed us to bring both the particle absorption and breathability closer to market products. The results have allowed to identify an optimal filter deposition time between 1 and 2 minutes (Figure 5). Since these thin filters are difficult to handle, the electrospinning process was adapted to deposit the filter directly on the outer layer of sheets of commercial facemasks (spun-bound non-woven polypropylene, supplied by EPSA-Swiss). Therefore, in this case the measurements of all electrospun and reference samples are made on an outer layer / filter composite. To evaluate the filter performance, we performed a differential measurement with the outer layer only. The pristine outer layer substrate was tested individually and its influence on the breathability and filtration was insignificant compared to the filter. Figure 5a shows the pressure drop of our filters, based on polymer **7** in comparison with commercial mask references. As expected, the pressure drops increase with filter thickness, for all the flow rates tested. The pressure drops for deposition times between 1-2 minutes correspond to the values of the commercial filters. In either case, our filters are within the above-mentioned norm. The particle capture efficiency of the filters is shown in Figure 5b. The filtration performance with a 2 min deposition time is better than our reference commercial filter, while the filter fabricated within 1 min exhibits a lower capture efficiency for all particle sizes except the largest (Figure 5b). Importantly, compared to our reference commercial filter, the electrospun PEA-polymer **7** filter with 2 min deposition time shows a significantly higher absorption rate for both 1 and 3 µm particles while respecting the breathability norm (we also refer to the discussion for this point).

In summary, by controlling the thickness of the filters through deposition time and using the optimum electrospinning deposition parameters, we demonstrated performances comparable with commercial facemasks and with the norms evaluated.

Nanoindentation tests have been performed on selected PEA-polymer **7** filters with different deposition time and the two reference facemasks. These tests did not show any significant dependence of the mechanical stiffness on the electrospinning deposition time (not shown), but the electrospun fibres exhibited an elastic modulus twice as large as the commercial PP filters (see details and Figure S5 in SI), highlighting again the tuneability and excellent mechanical properties of our poly(ester amide) polymer.

## Meltspinning

Electrospinning is a solution of choice for the fabrication of critical, high-performance filters as shown above and elsewhere in the literature[38]. However, due to its limited throughput, it is not the most appropriate for the fabrication of the outer layer. In this case, more conventional approaches such as melt-spinning are typically favoured, especially since the fibres thus created have a lower constraint in terms of performance efficiency. Therefore, we decided to evaluate the processability of our selected polymers to create non-woven fabric using melt-spinning.

Melt-spinning tests have been successfully applied with polymer **4** as demonstrated by continuous fibers fabrication (Figure 6b). The polymer pellets were maintained at a controlled temperature inside the dispenser head and continuous fibers were created and collected by the rotating cylindrical collector (Figure 6c). Due to the speed limitations of the rotating collector (700 rpm) the smallest fiber dimensions that could be achieved was 34 µm.

Polymer **4** was the only PEA grade that could be used for melt-spinning as no continuous fibers were generated using the two other selected grades. Viscosity versus temperature scans of these three grades were performed (Figure 6a) to investigate the reason behind the processability of the polymers.

The figure highlights two distinct behaviours that differentiate polymer **4** from polymers **1** and **7**. The slope of the viscosity around the melting temperature, and the final viscosity value reached in the molten state. When heated up to the melting point, polymer **4** shows a lower viscosity drop with temperature compared to the other two other PEA grades. Furthermore, over a temperature range of ΔT>16K the viscosity remains above 3000 Pa·s in the molten state and is much greater for polymer **4** than for the other two grades. This analysis therefore confirms why polymer **4** is compatible with melt-spinning while the other two are not. This is because in the latter cases, small temperature fluctuations can lead to a rupture of the fibre flow.

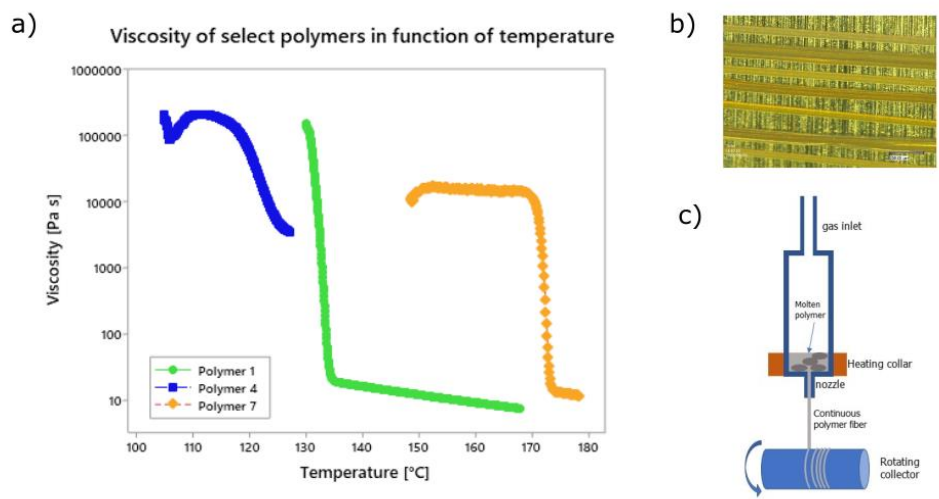

**Figure 6.** *PEA grade optimised for melt-spinning process. a) Viscosity versus temperature for the three PEA grades around the melting point. Polymer 4 shows a smoother variation of viscosity around the melting point in comparison to polymer grades 1 and 7. This explains why the grade 4 is less sensible for temperature fluctuations and better fits for melt-spinning. b) Optical micrograph of continuous fibres as obtained by melt-spinning the polymer 4 c) Principle of the melt-spinning equipment. The molten polymer is pushed through a nozzle and collected by a rotating cylinder.*

## Discussion

In this paper we present and fully characterise novel PEA-based biodegradable facemasks components made. Several biosourced PEA grades were synthesised, fine-tuned and evaluated for processing into non-woven fibres. Normalised biodegradation tests were performed on three PEA grades candidates that passed a processability screening test. Two polymers were fully degraded in less than 35 days, including one that degraded as quickly as cellulose, that is 20 days.

We then developed electrospun filters that were tested against filtration efficacy and breathability, using a test bench developed for the occasion. We systematically tested the filtration performance with particles between 0.3 and 10 micrometer size. Our optimised ultra-thin high-performance filters demonstrate filtration efficiency and breathability comparable to commercial facemasks.

The filtration efficacy and breathability of the filters were benchmarked against commercial filters. Disposable facemasks of Type I need to filter 95% of 3 µm particles (norm EN 14683+AC:2019). However, during the COVID-19 pandemic several Swiss hospitals and medical institutes have requested to add a 70% filtration of 1 µm particle criteria [28]. Commercial filter used as reference herein have been validated using both criteria and the mask used as reference for the filtration test has shown a 95% absorption of 1 µm particles. Furthermore, both commercial masks demonstrated a breathability that was approximately twice better than the normed target value (40 Pa/cm$^2$ according to norm EN 14683+AC:2019). Since the performance of our optimized filters are comparable to that of the two reference filters, we conclude that they may respond to the requirements of a normed test.

One further key outcome of our study is the demonstration that the filter material can directly be electrospun on the outer layer of a disposable mask. This ensures a tight contact between the layers and importantly simplifies the production of multi-layered systems. In addition, to evaluate the possibility to fabricate the outer layers of the facemasks, we developed a melt-spinning rig capable of continuous fibre fabrication on one of the polymers.

Pandit et al.[19] also claim that biodegradable materials will not only reduce waste but also increase wearing comfort and skin friendliness. Several authors report studies where biosourced and biodegradable filters have been made by electro-spinning, for example based on gluten blended PVA[39] or carbon blended gluten nanofiber[40], while other authors propose biopolymers like PVA and PLA[41] for electro-spinning nanofilters. To the best of our knowledge there are no biodegradable face masks based on biosourced PEA. PEA is an interesting class of polymers that combines good mechanical properties, biodegradability and importantly that is amenable to seamless modifications to fine tune their properties as shown herein.

One open point of this research pertains to the solvent used for electro-spinning, HFIP, that is not a green solvent. Despite comprehensive tests with other solvents like methanol and ethanol and mixtures of HFIP with green solvents the presented electro-spinning results have only been achieved using pure HFIP.

Overall, our results pave the way for the development of high-performance biodegradable facemasks based on biosourced PEA. Future studies, outside of the scope of this manuscript, will address the ideal combination of melt-spinning and electro-spinning process to fabricate an entire facemask. The potential final costs of a disposable biodegradable facemask will also be evaluated and factors like the upscaled PEA synthesis and a production including an electrospinning step will be considered.

# Materials and methods
## Synthesis of biodegradable Polyesteramide polymer

**Polymer Synthesis & Characterization.**

General Information: Dimethyl 2,5-furandicarboxylate (DMFD) was purchased from Apollo Scientific, 1,4-butanediol; 1,6-hexanediol; 1,10-decanediol; DBTO were purchased from Acros Organics. Diethylether; absolute alcohol with 5% IPA; tetrahydrofuran with BHT stabilizer; methanol, 99% were purchased from Thommen-Furler AG and used without any further purification; titanium (IV) butoxide was purchased from Fluorochem and dissolved in toluene at the desired concentration prior to use.

NMR spectra were recorded with a Bruker 300 Ultrashield spectrometer and referenced against the chemical shift of the residual protio-solvent peak ($CDCl_3$: 7.26 ppm; DMSO-d6: 2.50 ppm, $D_2O$: 4.79 ppm) for 1H NMR and the deuterated solvent peak ($CDCl_3$: 77 ppm; DMSO-d6: 40 ppm) for 13C NMR measurements.

Infrared spectra were recorded on a Bruker ALPHA in absorption mode between 4000 $cm^{-1}$ and 400 $cm^{-1}$ with a resolution of 4 $cm^{-1}$. Samples were analysed directly on the diamond crystal without further preparation (not shown).

DSC measurements were carried with a Mettler DSC 821e. The analyses were conducted under nitrogen in Al 40 µL crucibles with a heating and cooling rate of 10°C/min. Method: -70°C to 200°C, 1 min annealing, 200°C to -70°C, 1 min annealing, -70°C to 200°C.

A TGA/SDTA851e (Mettler Toledo) instrument was used to study the thermal stability of the synthesized polymers. To this purpose, 5-10 mg of polymers were placed in a standard aluminium pan and heated under nitrogen from 30°C to 800°C at a heating rate of 10°C/min. The first indicator is the temperature for which the weight loss is equal to 5% ($T_{d,5\%}$).

GPC measurements were performed on a Waters 1260 infinity pump, a 1260 Infinity II Refractive Index Detector, a 1260 Infinity II Multisampler, an Acquity APC XT 45,1.7µm column an Acquity APC XT 125, 2.5 µm followed by an Acquity APC XT 200, 2.5 µm columns in series at 30 °C. A 10 mM solution of Sodium triacetate in Hexafluoroisopropanol was used as eluent at a flow rate of 0.3 mL/min. The molecular weights were calibrated with PMMA standards on a range between 600 and 2 200 000 Da (PSS Polymer Standards Service, Mainz, Germany).

For the test of polymer solubility, a defined amount of polymer and solvent was charged in a vial and left stirring until dissolution. Solubility was evaluated visually.

**Synthesis of Poly(ester amide)s**

The **6,4-bisamidediol building block** was synthesized as reported in references [42,43].

1H-NMR signals of 6,4-bisamidediol (300 MHz; DMSO-d6): δ 7.74 (t, J=5.6 Hz, 1H), 4.39-4.30 (m, 1H), 3.37 (dd, J=6.4; 4.7 Hz, 2H), 3.00 (q, J=6.1 Hz, 2H), 2.03 (t, J=7.4 Hz, 2H), 1.58-1.15 (m, 3H).

Polymers were synthesized following the same procedure that comprises a first step where transesterification of DMFD or dimethyl adipate occurs with the consequent removal of methanol and a second polycondensation step where the polymer chains grow, and the removal of the diol takes place. An example the synthesis polymer **2** with DMFD and 1,4-butanediol to form a 50% hard segment is reported.[43]

In a 500 mL three-necked round-bottom flask mounted with a distilling bridge and a helical stirrer connected to the system via a magnetic coupling, 58.18 g of DMFD (321.19 mmol, 1 eq.), 50.93 g of building block 6,4-bis-amidediol (97.5% purity, 157.75 mmol, 0.49 eq.), 14.47 g of 1,4-butandiol (160.56 mmol, 0.5 eq.) were introduced and 3 cycles of argon/vacuum were made to ensure inert atmosphere. The flask was heated to 190°C using an aluminium heating block (DrySyn) under argon atmosphere. Once the reactants melted, the stirring was enabled at 200 rpm and 4 mL of a 30 mg/mL stock solution in toluene of titanium tert-butoxide (catalyst, 180 mg, 0.52 mmol) were added through a septum. During the esterification a stream of Argon was purged through the reactor to remove methanol and toluene efficiently. Once the distillation ended the distillation collector was emptied, dried and re-connected. Then, 4 mL of a 30 mg/mL stock solution in toluene of titanium tert-butoxide (catalyst, 180 mg, 0.52 mmol) were added through a septum.

The temperature was increased to 205°C, the pressure was reduced to 0.02 mbar for 1.5 hours with a high vacuum pump. After this time the pressure was increased with Argon up to atmospheric pressure and 4 mL of a 30 mg/mL stock solution in toluene of titanium tert-butoxide (catalyst, 180 mg, 0.52 mmol) were added through a septum. After reducing the pressure again, the solution was kept 1.5 h at 205°C. Then, the temperature was increased to 210°C for 1.5 h and finally to 215°C for 1 h. After this time the formed polymer (lightly brown viscous liquid) was cast on a metal plate to allow solidification.

### Synthesis of 2,2'-bis(2-oxazoline)

2,2'-bis(2-oxazoline) was synthesized based on the procedures reported by H. Wenker[31] and in the patent WO2012066051A2[44].

Diethyl oxalate (14.6 g, 0.1 mol, 1 eq.) dissolved in 15 mL of ethanol is added for 1 hour to a cooled mixture composed of 2-chloroethylamine hydrochloride (23.2 g, 0.2 mol, 2 eq.) and potassium hydroxide (85%, 13.2 g, 0.2 mol, 2 eq.) dissolved in 20 mL of deionized water. The mixture temperature is kept below 20°C with an ice bath. At the end of the addition, the mixture is stirred for an additional hour at room temperature. The white precipitate is isolated by vacuum filtration, suspended in 40 mL of deionized water, and stirred for 15 minutes. The suspension is filtered again by vacuum filtration and washed with 15 mL of ethanol. The powder is dried in a vacuum oven at 80°C and 50 mbar. 16.92 g of N,N'-bis(2-chloroethyl)oxamide as a fine white powder is obtained (yield: 79%).

m.p.: 203°C (203°C, Lit.(35))

1H NMR (300 MHz, DMSO-d6) δ 8.95 (t, J = 6.3 Hz, 2H, N-H), 3.70 (t, J = 6.3 Hz, 4H, Cl-CH2), 3.49 (q, J = 6.2 Hz, 4H, N-CH2)

13C NMR (75 MHz, DMSO-d6) δ 160.4 (C=O), 43.0 (C-Cl), 41.3 (C-NH)

IR: 3291s, 3067w, 2961w, 2934w, 1655s, 1534s, 1440s, 1362w, 1311m, 1246s, 1185m, 1055m, 933w, 860w, 760m, 652m, 546m.

N,N'-bis(2-chloroethyl)oxamide (13.42 g, 63 mmol, 1 eq.) is suspended in 50 mL of methanol containing 8.32 g of potassium hydroxide (85%, 126 mmol, 2 eq.). The mixture is heated to reflux for one hour. The resulting suspension is filtered by vacuum filtration at 50°C. The filtrate is concentrated at 50°C and 200 mbar. When around 40 mL of methanol is removed, the precipitate is filtered off under vacuum and washed with a small volume (about 5 mL) of cold methanol. A second concentration and filtration step is performed on the resulting filtrate with the same parameters. The powder obtained from these two concentrations is dried in a vacuum oven at 60°C and 100 mbar. 7.05 g of 2,2'-bis(2-oxazoline) as a white crystalline powder is obtained. (Yield: 80%).

m.p.: 213°C (213°C, Lit.(35))

1H NMR (300 MHz, D2O) δ 4.49 (t, J = 9.9 Hz, 4H, O-CH2), 4.00 (t, J = 9.9 Hz, 4H, N-CH2).

13C NMR (75 MHz, D2O) δ 156.09 (C=O), 69.24 (O-CH2), 53.73 (N-CH2)

IR: 3291w, 2940w, 2872w, 1655w, 1617s, 1538w, 1473w, 1385w, 1348w, 1290w, 1253w, 1188w, 1105s, 971m, 913s, 868m, 725w, 653w, 569m.

**2,2'-bis(2-oxazoline)-Based Poly(ester amide) synthesis**

Bulk polymerizations of PEAs were carried out according to the procedure reported by Wilsens[45].

Sebacic acid (6.06 g, 30 mmol, 1 eq.) is mixed with 2,2'-bis(2-oxazoline) (4.62 g, 33 mmol, 1.1 eq.) and Irganox 1330 (0.1 g, 1 wt%). The mixture is mixed and heated under nitrogen atmosphere to 195°C for two hours, before the viscous polymer is discharged on a Teflon foil.

**Aerobic biodegradation of polymers**

Polymers **1**, **4** and **7** were grounded with Ultra Centrifugal Mill ZM 200 from Retch at a size lower than 800 μm and sent to OWS[46].

The controlled composting biodegradation test is an optimized simulation of an intensive aerobic composting process where the biodegradability of a test item under dry, aerobic conditions is determined. The inoculum consists of stabilized and mature compost derived from the organic fraction of municipal solid waste. The test item is mixed with the inoculum and introduced into static reactor vessels where it is intensively composted under optimum oxygen, temperature and moisture conditions. During the aerobic biodegradation of organic materials, a mixture of gases (principally carbon dioxide and water) are the final decomposition products while part of the organic material will be assimilated for cell growth. The carbon dioxide production is continuously monitored and integrated to determine the carbon dioxide production rate and the cumulative carbon dioxide production. After determining the carbon content of the test item, the percentage of biodegradation can be calculated as the percentage of solid carbon of the test item, which has been converted to gaseous, mineral C under the form of $CO_2$.

The tests were performed under the norm: ISO 14855-1 Determination of the ultimate aerobic biodegradability of plastic materials under controlled composting conditions – Method by analysis of evolved carbon dioxide (2012), but in singular instead of triplicate.

## Fibre fabrication through electrospinning of PEA

The Genvolt electrospinning starter kit, composed of a high voltage power supplier (up to 30 kV) and a syringe pump, was used for early screening. Glass syringes with a metallic needle and a rotating collector were used for the experiments.

Polymers were dissolved in a vial at a given concentration (see Table S1 in SI, for details) and left stirring with a magnetic stirrer overnight at room temperature. The polymer solution was then filtered on a 0.22 μm filter and loaded into a 10 mL glass syringe. The syringe was put on the syringe pump and the voltage connected to the needle. The flow of the syringe pump and the collector speed (rpm) were selected, and the tension was then applied. The tension was slowly increased until the apparition of fibres from the tip of the needle. After the process was finished the aluminium foil was removed from the collector and the tissue composed of polymer fibres was further used for SEM analysis.

Final electrospinning experiments on selected polymers were realised with a 4Spin® device using a 21-gauge needle as the single emitter and a fixed rectangular collector. The setup is represented in Figure 1a PEA grades were dissolved in hexafluoroisopropanol (HFIP) with concentrations ranging between 8 wt.% and 12.5 wt.% for the selected polymers. Fibres were deposited on aluminium foil or commercial mask outer layer, depending on the condition evaluated, on the collector side. Voltage, working distance and flowrate were optimised through experimental design, as detailed in supplementary information and table S2.

Using the results from the experimental design, selected solutions were electro-spun with an emitter-collector distance of 13 cm, a voltage of 17 kV and a feed rate of 30 µL/min. Samples for filter performance characterisation were made by cutting 50 mm wide disks out of the fabric at the point where the fibres were the thickest, as assessed visually. Deposition time and polymer choice were the only variable for filter fabrication and three filters were made for each parameter setup.

## Assessment of fibre quality through scanning electron microscopy

Quality of electrospun fibres was assessed visually using a JEOL JSM-6400 scanning electron microscope on metalized samples. Even spreads of long uniform fibres were judged of good quality while the presence of polymer beads indicated electro-spraying. Samples of the experimental design were graded from 1 to 6 (Table S2 in SI) and sorted onto a heatmap to better represent the influence of the variable parameters (Figure 4 and Figure S3).

## Measurement of layer thickness through confocal microscopy

The thickness of layers deposited through electrospinning follows a gaussian distribution with the centre of the stream producing a wider fabric than on the sides. In order to get a measurement of thickness, it is important to account for that variability so that different measurements can be compared. This was done by depositing fibres over a silicon wafer partially covered by peelable masks in a fixed position. A confocal microscope (Sensofar S neox) was then used to measure the height differences over a line between the areas where the mask was peeled off. To ascertain the relationship between layer thickness and deposition time, fibres deposited over 30, 60, 90 and 120 min were prepared for PEA 50% and up to 90 min for PEA 25%.

## Nanomechanical tests of electrospun PEA fibers and commercial masks

Nanoindentation tests of individual filter filaments have been performed with an Ultra nanoindenter UNHT (Anton-Paar) equipped with a Berkovich tip. This local probe method is explained in more detail in many publications, for example[47]. Briefly, the diamond tip is loaded into the sample and the force is recorded as a function of the displacement. During the loading phase the material deforms plastically and elastically whereby both contributions cannot be distinguished. During the unloading phase the material recovers elastically, which allows to determine the elastic modulus and hardness. For these nanomechanical tests the filter samples have been embedded in PMMA and polished. Indentation tests have been performed on individual fibers of selected electrospun PEA filters and of the reference commercial facemasks (see below). For each sample 15 indentation tests have been performed with a maximum load of 200 µN. A linear loading rate of 200 µN/min was applied, followed by 10 s break at maximum load and an unloading applying a rate of 200 µN/min. The local elastic modulus and hardness of the fiber material were determined using a Poisson's ratio of 0.3.

## Rheogical analysis of the PEA polymers

Rheological tests have been performed with an Anton Paar MCR702 TwinDrive rotational rheometer. Viscosity versus temperature scans have been carried out for selected PEA grades under a nitrogen atmosphere to avoid oxidation. The temperature interval has been defined based on the DSC scans (see Table 2). The measured range covered the melting point of the individual PEA and the temperature has been varied at 1°C/min applying an oscillation frequency of 1 Hz and an imposed 0.8%-1% relative deformation.

### Custom made filter test bench for breathability and absorption test
A filter test bench to measure pressure drop and microparticles absorption rate was conceived and realized in-house. The custom-made bench architecture and working principles are presented in detail in the results section. Teflon microparticles (Polytetrafluoroethylene PTFE Powder; CAS: 9002-84-0) with particle diameters ranging from 0.3 to 10 µm were used. The number of particles passing through the filters were counted using a particle counter (HPPC6 Particle Counter Plus 8306; Connect 2 Cleanrooms Ltd, Lancaster, UK) with an integrated pump and detection range of 0.3, 0.5 1, 3, 5 and 10 µm particles. Commercial disposable facemasks, Einwegmaske "Facemask", PP, EN 14683:2019+AC:2019 and Aspop Einwegmaske PP en 14683 Type IIR, validated in terms of particle absorption and breathability by Kassensturz (October 2020), were used as reference.

### Melt spinning tests
A melt-spinning set-up, allowing for solvent-free fabrication of continuous polymer fibres, was fabricated. It consisted of an extrusion head (Fig. 1b and Fig. S6 in supplementary materials) and a rotating cylindrical collector with a maximum speed of 700 rpm. The polymer is molten in the head body, via a heating collar reaching up to 400°C and pushed through the nozzle using compressed air.

## Acknowledgements

We would like to acknowledge funding from HES-SO (Projet Libre, Public Mask). We are grateful to C. Csefalvay for scanning electron microscopy and confocal microscopy measurements. We are grateful to M. Choain for supplying EPSA Swiss mask components.

# Supplementary Information

# Tuneable and biodegradable poly(ester amide)s for disposable facemasks

Esteban Alvarez Seoane, Alessandro Cattaneo, Tatiana Nogueira Matos, Laure Jeandupeux, Gianni Fiorucci, Maryam Tizgadam, Kelly Tran, Pierre-Louis Sciboz, Luce Albergati, Jérôme Charmet, Roger Marti, Stefan Hengsberger

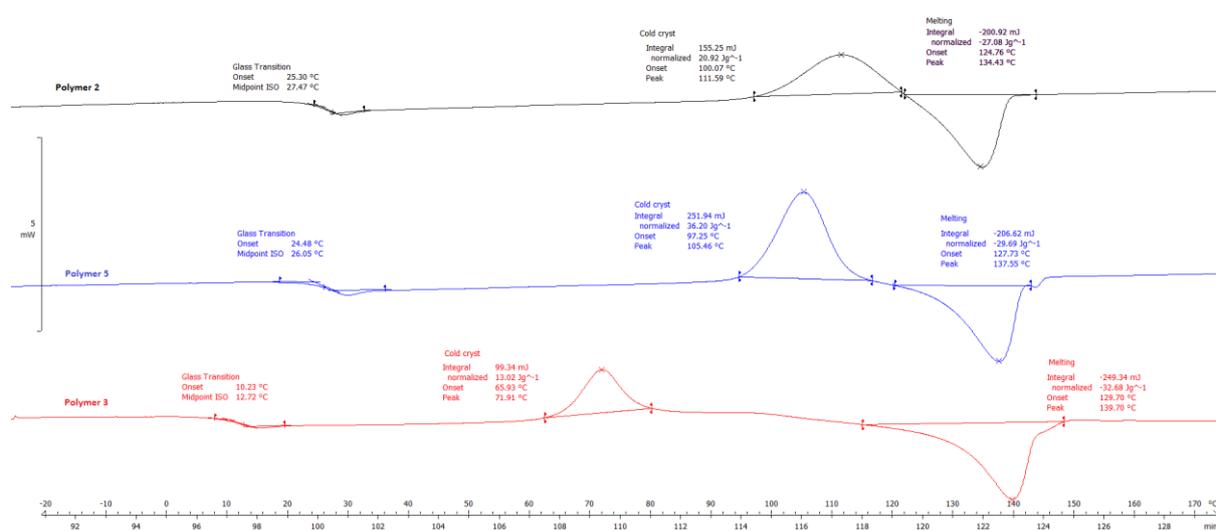

**Figure S1.** Differential Scanning Calorimetry thermograms of PEA 50-50 ester amide ratio synthesized from different diols.

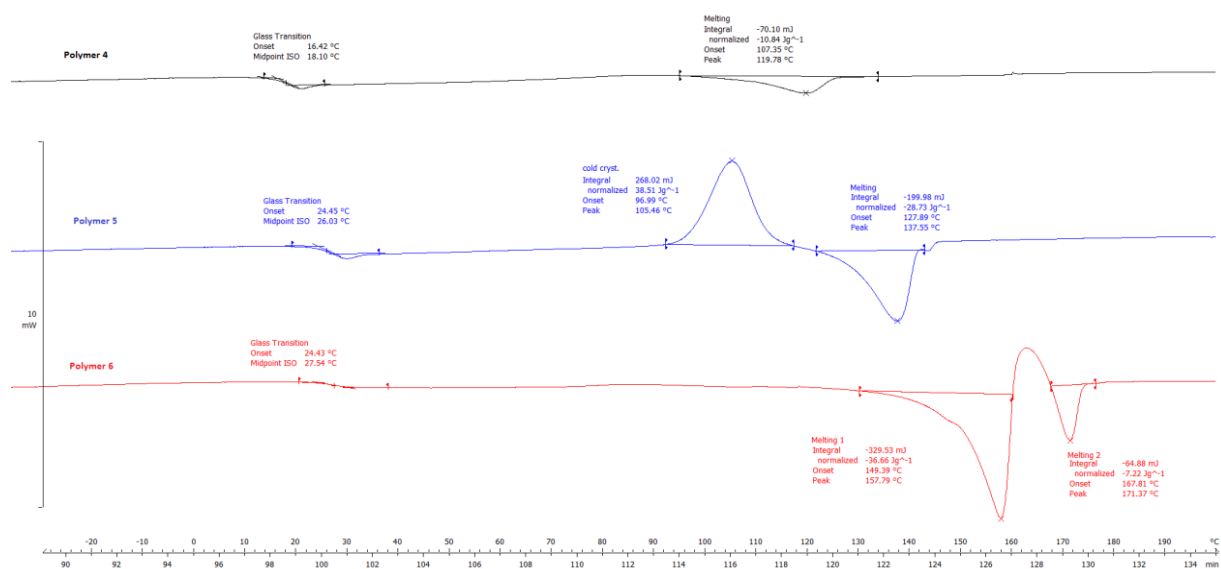

**Figure S2.** Differential Scanning Calorimetry thermograms of different PEA grades synthesized from 1,6-hexanediol.

Table S1. Polymer solubility was tested in different solvents.

| Polymer | Solvent 1 | Solvent 2 | Conc. [%wt.] | Observations |
|---|---|---|---|---|
| 2 | Hexafluoro-2-propanol | - | 7.5 | Soluble |
| 3 | Hexafluoro-2-propanol | - | 7.5 | Soluble |
| 2 | Chloroform | Methanol | 6.25 | Soluble |
| 3 | Chloroform | Methanol | 6.25 | Soluble |
| 7 | Chloroform | Methanol | 6.25 | Not soluble |
| 2 | N-Methylmorpholine N-oxide | - | 5 | Not soluble |
| 3 | N-Methylmorpholine N-oxide | - | 5 | Not soluble |
| 7 | N-Methylmorpholine N-oxide | - | 5 | Not soluble |
| 2 | Chloroform | Benzyl alcohol | 6.7 | Soluble |
| 3 | Chloroform | Benzyl alcohol | 6.7 | Soluble |
| 7 | Chloroform | Benzyl alcohol | 5.0 | Not soluble |
| 2 | Methanol | Benzyl alcohol | 6.7 | Soluble |
| 3 | Methanol | Benzyl alcohol | 6.7 | Heating necessary |
| 7 | Methanol | Benzyl alcohol | 6.7 | Not soluble |
| 2 | Chloroform | Phenyl ethanol | 6.7 | Soluble |
| 3 | Chloroform | Phenyl ethanol | 6.7 | Soluble |
| 7 | Chloroform | Phenyl ethanol | 6.7 | Not soluble |
| 2 | Methanol | Phenyl ethanol | 6.7 | Soluble |
| 3 | Methanol | Phenyl ethanol | 6.7 | White suspension after heating |
| 7 | Methanol | Phenyl ethanol | 6.7 | Not soluble |
| 2 | Chloroform | Ethanol | 6.7 | Soluble |
| 3 | Chloroform | Ethanol | 6.7 | Soluble |
| 7 | Chloroform | Ethanol | 6.7 | Not soluble |
| 2 | Ethanol | Benzyl alcohol | 6.7 | Soluble |
| 3 | Ethanol | Benzyl alcohol | 6.7 | White suspension after heating |
| 7 | Ethanol | Benzyl alcohol | 6.7 | Not soluble |
| 2 | Ethanol | Phenyl ethanol | 6.7 | Soluble |
| 3 | Ethanol | Phenyl ethanol | 6.7 | White suspension after heating |
| 7 | Ethanol | Phenyl ethanol | 6.7 | Not soluble |
| 2 | Chloroform | Formic acid | - | solvents forms 2 phases |

| Polymer | Solvent 1 | Solvent 2 | Conc. [%wt.] | Observations |
|---|---|---|---|---|
| 3 | Chloroform | Formic acid | - | solvents forms 2 phases |
| 7 | Chloroform | Formic acid | - | solvents forms 2 phases |
| 2 | Dichloromethane | Methanol | 6.7 | Soluble |
| 3 | Dichloromethane | Methanol | 6.7 | Heating necessary |
| 7 | Dichloromethane | Methanol | 6.7 | Not soluble |
| 2 | Dichloromethane | Ethanol | 6.7 | Soluble |
| 3 | Dichloromethane | Ethanol | 6.7 | Heating necessary |
| 7 | Dichloromethane | Ethanol | 6.7 | Not soluble |
| 2 | Dichloromethane | Benzyl alcohol | 6.7 | Soluble |
| 3 | Dichloromethane | Benzyl alcohol | 6.7 | Soluble |
| 7 | Dichloromethane | Benzyl alcohol | 6.7 | Not soluble |
| 2 | Dichloromethane | Phenyl ethanol | 6.7 | Soluble |
| 3 | Dichloromethane | Phenyl ethanol | 6.7 | Soluble |
| 7 | Dichloromethane | Phenyl ethanol | 6.7 | Not soluble |
| 2 | Methanol | Dimethyl carbonate | 6.7 | Soluble |
| 3 | Methanol | Dimethyl carbonate | 6.7 | Not soluble |
| 7 | Methanol | Dimethyl carbonate | 6.7 | Not soluble |
| 2 | Ethanol | Dimethyl carbonate | 6.7 | Soluble |
| 3 | Ethanol | Dimethyl carbonate | 6.7 | Not soluble |
| 7 | Ethanol | Dimethyl carbonate | 6.7 | Not soluble |
| 2 | Dichloromethane | Dimethyl carbonate | 6.7 | Soluble |
| 3 | Dichloromethane | Dimethyl carbonate | 6.7 | Not soluble |
| 7 | Dichloromethane | Dimethyl carbonate | 6.7 | Not soluble |
| 2 | Chloroform | Dimethyl carbonate | 6.7 | Soluble |
| 3 | Chloroform | Dimethyl carbonate | 6.7 | Soluble |
| 7 | Chloroform | Dimethyl carbonate | 6.7 | Not soluble |
| 5 | HFIP | - | 8.0 | Soluble |
| 5 | Chloroform | Methahol | 10.8 | Not soluble |
| 5 | Chloroform | Benzyl alcohol | 15.3 | Soluble |
| 5 | MeOH | Benzyl alcohol | 9.4 | Not soluble |
| 5 | Chloroform | Phenyl ethanol | 10.0 | Soluble |
| 5 | Chloroform | Ethanol | 9.4 | Soluble |
| 5 | Dichloromethane | Ethanol | 9.4 | Soluble |
| 5 | Dichloromethane | Methanol | 13.1 | Soluble |

| 5 | Dichloromethane | Benzyl alcohol | 13.1 | Soluble |
| --- | --- | --- | --- | --- |
| 5 | Dichloromethane | Phenyl ethanol | 12.5 | Soluble |
| 5 | Chloroform | Dimethyl carbonate | 6.3 | Not soluble |
| 4 | HFIP | - | 10.3 | Soluble |
| 4 | HFIP | - | 8 | Soluble |
| 4 | Chloroform | Methahol | 10.8 | Soluble |
| 4 | Chloroform | Benzyl alcohol | 9.4 | Soluble |
| 4 | Chloroform | Phenyl ethanol | 9.4 | Soluble |
| 4 | Chloroform | Ethanol | 9.4 | Soluble |
| 4 | Dichloromethane | Ethanol | 9.4 | Soluble |
| 4 | Dichloromethane | Methanol | 9.4 | Soluble |
| 4 | Dichloromethane | Benzyl alcohol | 9.4 | Soluble |
| 4 | Dichloromethane | Phenyl ethanol | 9.4 | Soluble |

### Design of experiment for electrospinning of PEA-polymer 7

SEM images were graded 1 to 6 according to the following criteria:

1. No fibre formation is observed, the process is purely in the electro spraying regime.
2. Electro spraying is the dominant regime, but polymer strands start appearing among the polymer beads.
3. A fibre network starts being visible, but fibres are of irregular shape and contain polymer beads.
4. Both polymer beads and fibres are present, but they can be distinguished from each other.
5. A fibre network is observed but the fibres are of varying diameter and aren't going straight.
6. A network of long, individual and mostly straight fibres is observed and polymer beads are very few in number

**Table S2**: *Design of Experiment PEA-polymer 7, 10wt.% in HFIP*

PEA 10wt% in HFIP
Sample n°1
40kV
17cm
40µl/min
Grade: 4

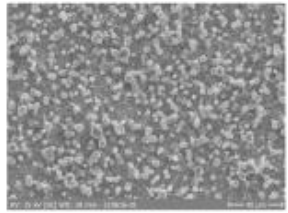 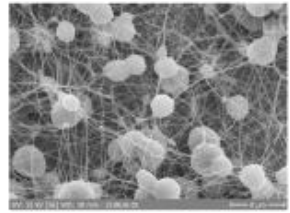 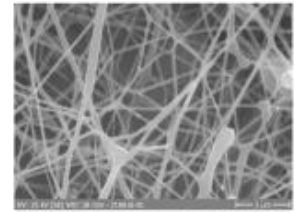

PEA 10wt% in HFIP
Sample n°2
40kV
10cm
120µl/min
Grade: 4

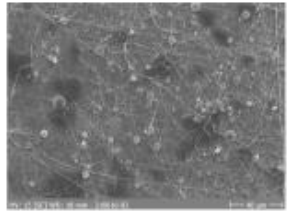 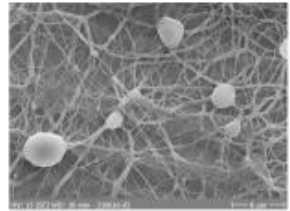 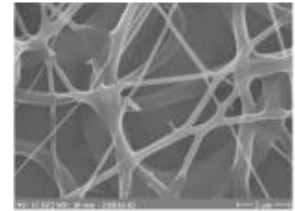

PEA 10wt% in HFIP
Sample n°3
26kV
17cm
120µl/min
Grade: 3

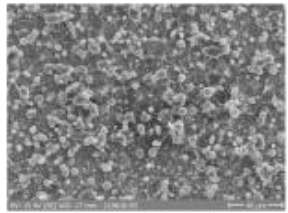 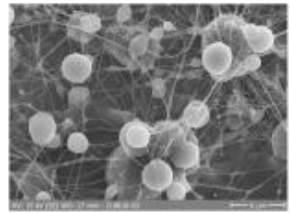 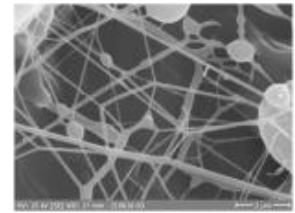

PEA 10wt% in HFIP
Sample n°4
26kV
25cm
120µl/min
Grade: 2

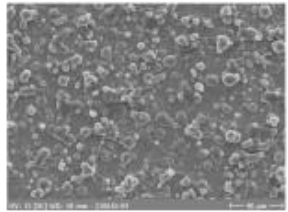 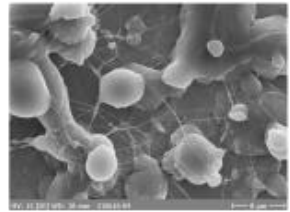 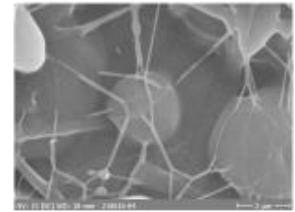

PEA 10wt% in HFIP
Sample n°9
26kV
17cm
10µl/min
Grade: 4

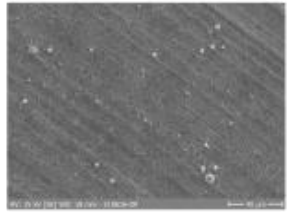 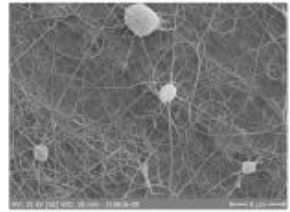 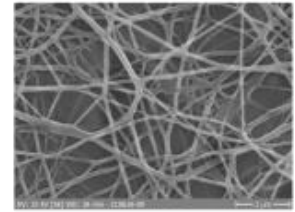

PEA 10wt% in HFIP
Sample n°10
12kV
17cm
40µl/min
Grade: 1

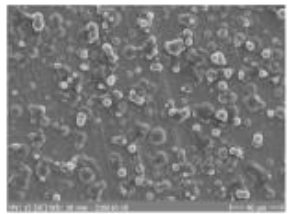 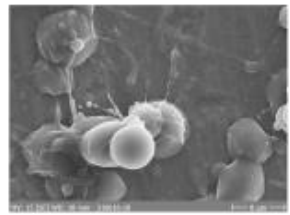 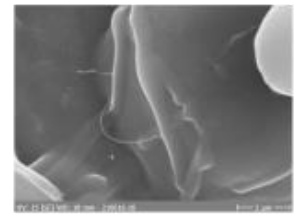

PEA 10wt% in HFIP
Sample n°11
26kV
10cm
10µl/min
Grade: 1

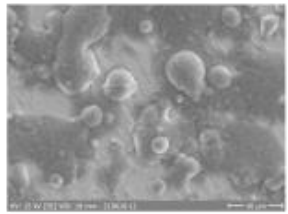 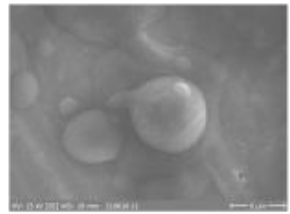 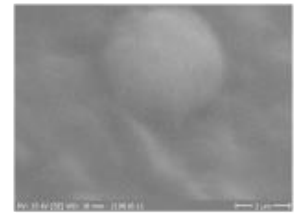

PEA 10wt% in HFIP
Sample n°13
26kV
25cm
40µl/min
Grade: 2

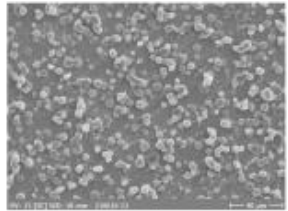 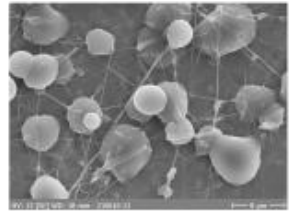 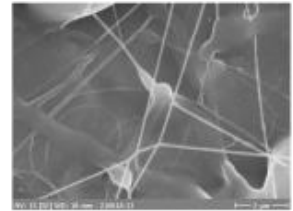

PEA 10wt% in HFIP
Sample n°14
12kV
10cm
10µl/min
Grade: 3

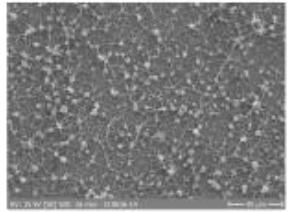 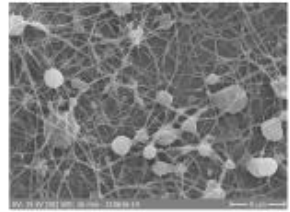 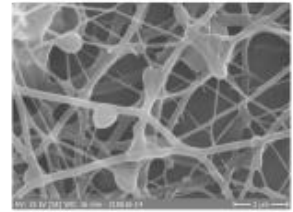

PEA 10wt% in HFIP
Sample n°15
40kV
25cm
120µl/min
Grade: 2

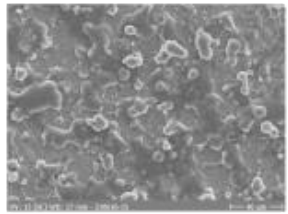 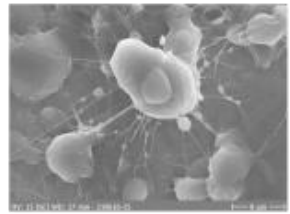 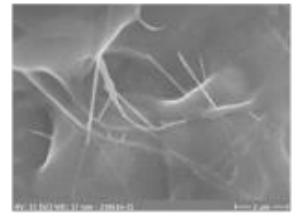

PEA 10wt% in HFIP
Sample n°16
12kV
25cm
10µl/min
Grade: 1

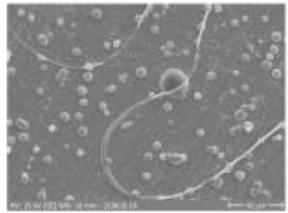 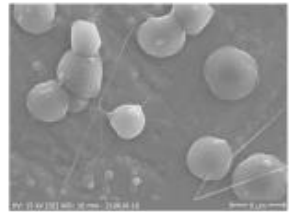 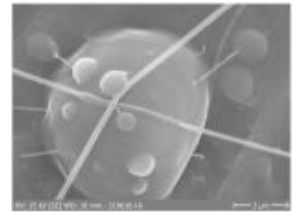

PEA 10wt% in HFIP
Sample n°17
26kV
25cm
10µl/min
Grade: 2

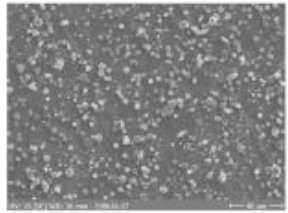 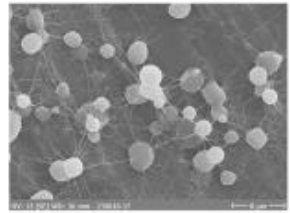 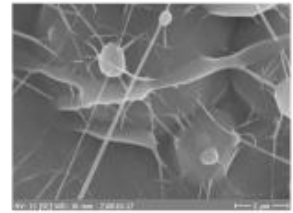

PEA 10wt% in HFIP
Sample n°18
12kV
10cm
40µl/min
Grade: 3

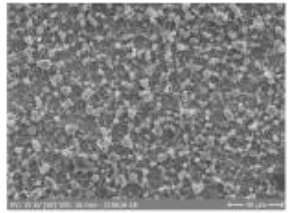 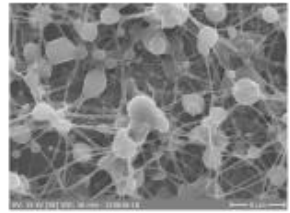 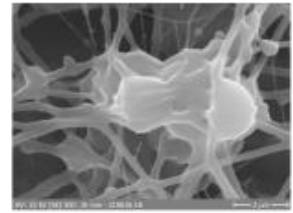

PEA 10wt% in HFIP
Sample n°19
12kV
10cm
120µl/min
Grade: 1

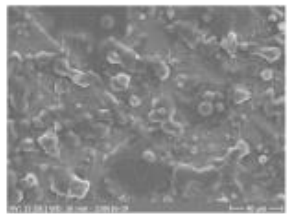 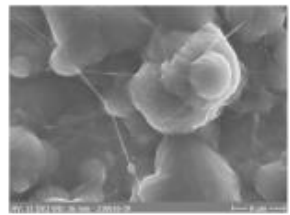 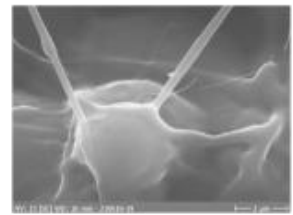

PEA 10wt% in HFIP
Sample n°20
12kV
17cm
10µl/min
Grade: 2

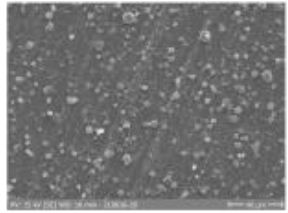 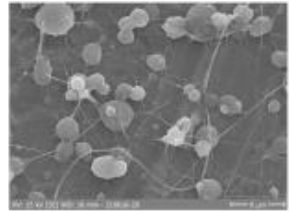 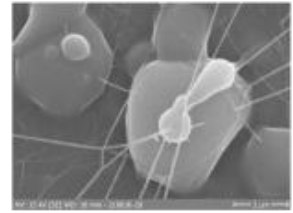

PEA 10wt% in HFIP
Sample n°21
26kV
17cm
40µl/min
Grade: 2

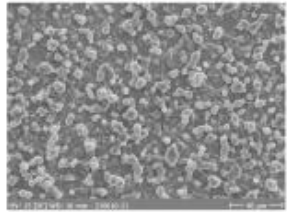 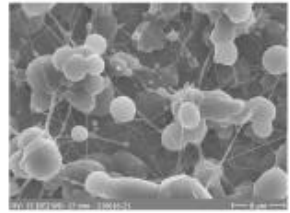 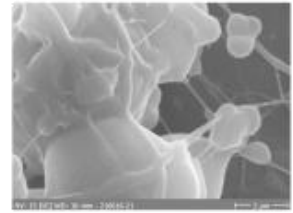

PEA 10wt% in HFIP
Sample n°23
40kV
25cm
40µl/min
Grade: 4

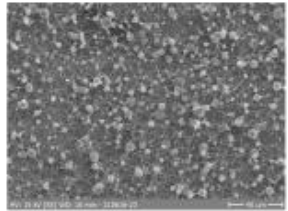 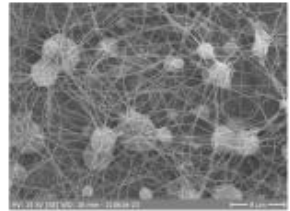 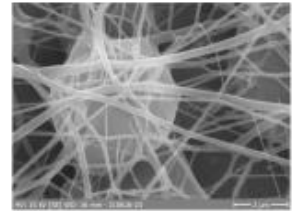

PEA 10wt% in HFIP
Sample n°24
40kV
17cm
120µl/min
Grade: 2

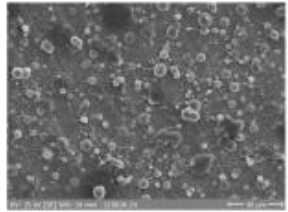 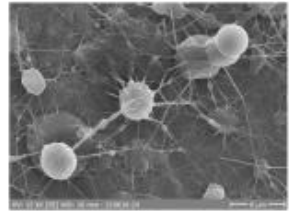 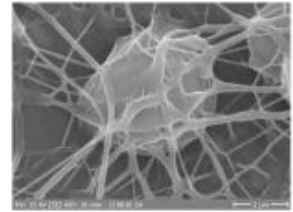

PEA 10wt% in HFIP
Sample n°25
40kV
25cm
10µl/min
Grade: 4

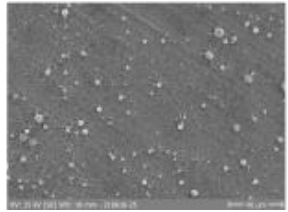 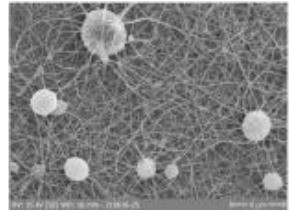 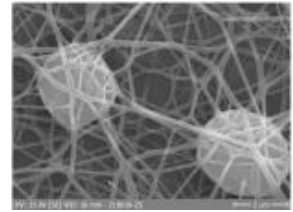

**Table S3:** *Design of Experiment PEA-polymer 7, 12.5wt.% in HFIP*

PEA 12.5wt% in HFIP
Sample n°1
10kV
20cm
10µl/min
Grade: 5

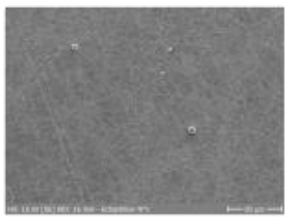 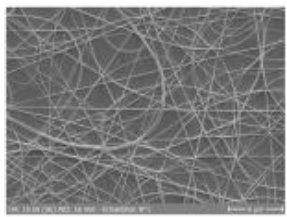

PEA 12.5wt% in HFIP
Sample n°2
25kV
10cm
10µl/min
Grade: 6

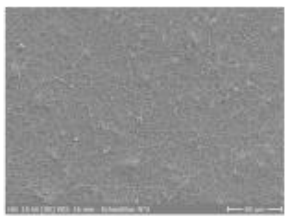 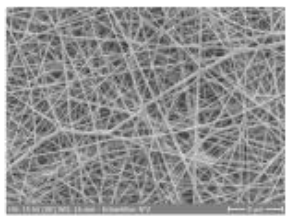 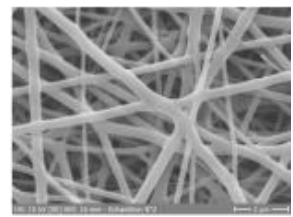

PEA 12.5wt% in HFIP
Sample n°3
17kV
20cm
30µl/min
Grade: 6

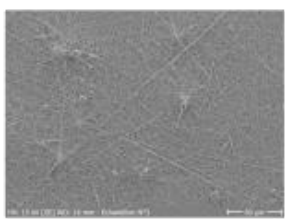 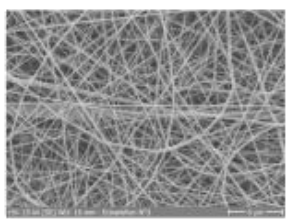 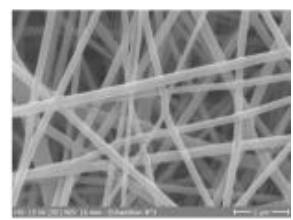

PEA 12.5wt% in HFIP
Sample n°4
25kV
13cm
30µl/min
Grade: 6

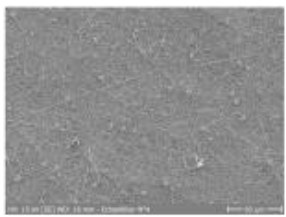 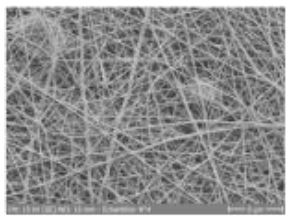 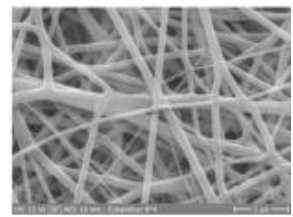

PEA 12.5wt% in HFIP
Sample n°5
17kV
13cm
50µl/min
Grade: 6

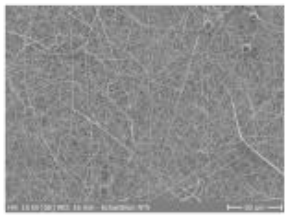 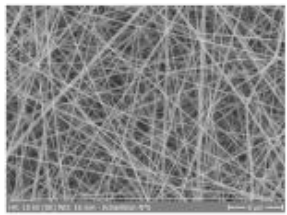 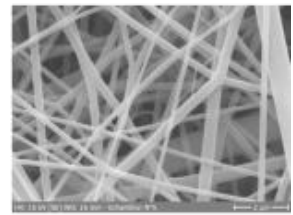

PEA 12.5wt% in HFIP
Sample n°6
17kV
20cm
70µl/min
Grade: 4

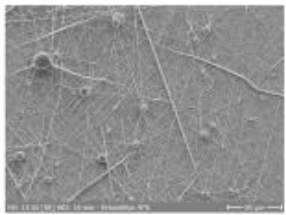 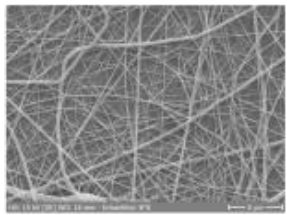

PEA 12.5wt% in HFIP
Sample n°7
25kV
13cm
70µl/min
Grade: 5

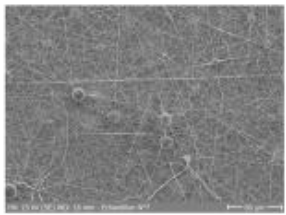 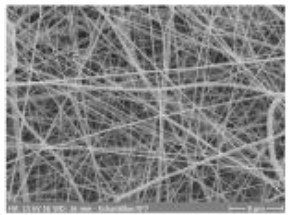 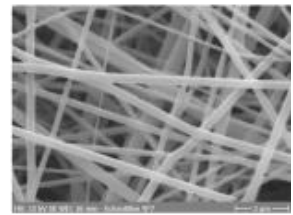

**Table S3:** *Design of Experiment PEA-polymer 7, 12.5wt.% in HFIP*

PEA 12.5wt% in HFIP
Sample n°8
10kV
20cm
90µl/min
Grade: 5

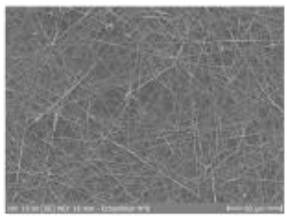 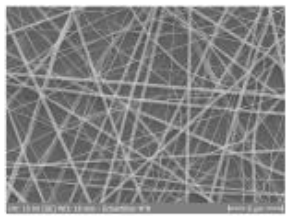 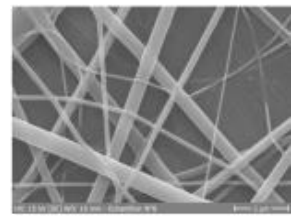

PEA 12.5wt% in HFIP
Sample n°9
25kV
10cm
90µl/min
Grade: 6

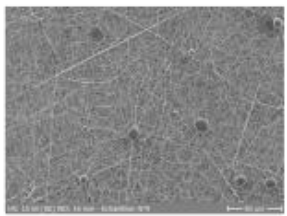 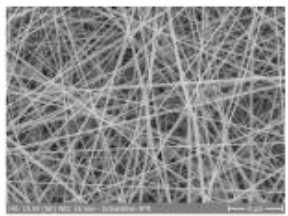 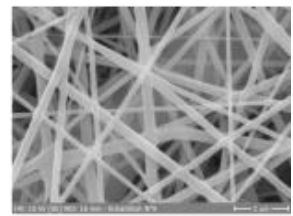

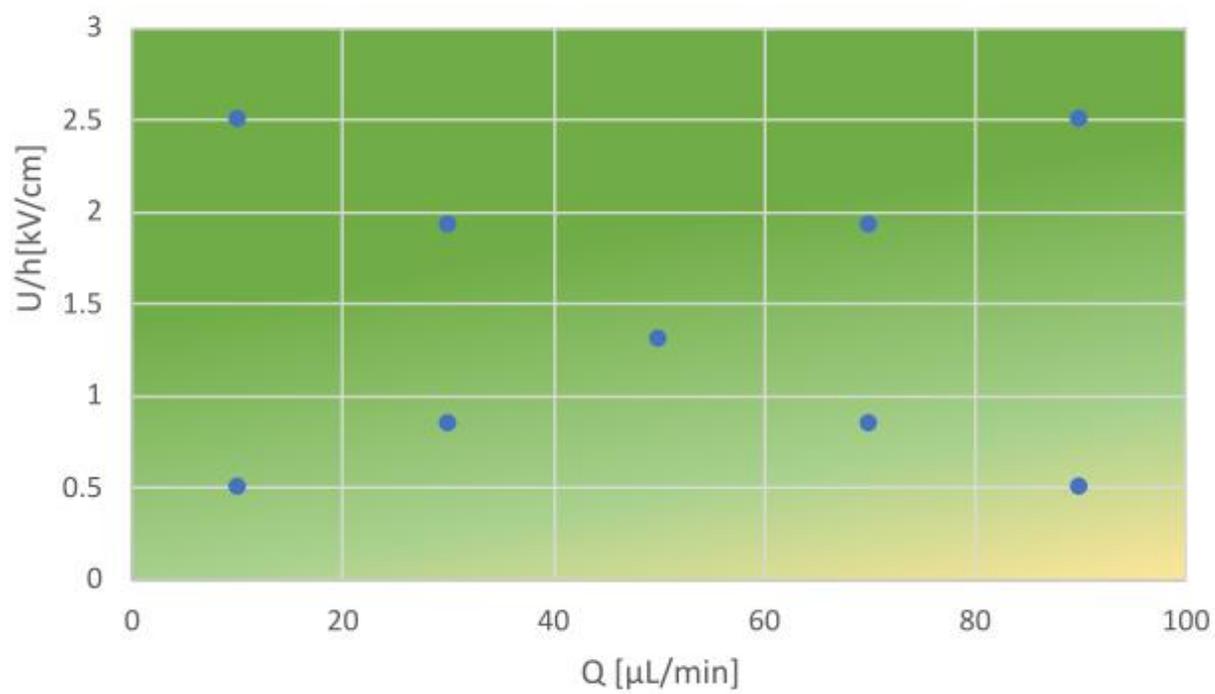

**Figure S3.** *Fibre quality heatmap as function of the flowrate and the ratio of voltage to distance to collector. Data for 12.5 wt.% solution of polymer **7** in HFIP.*

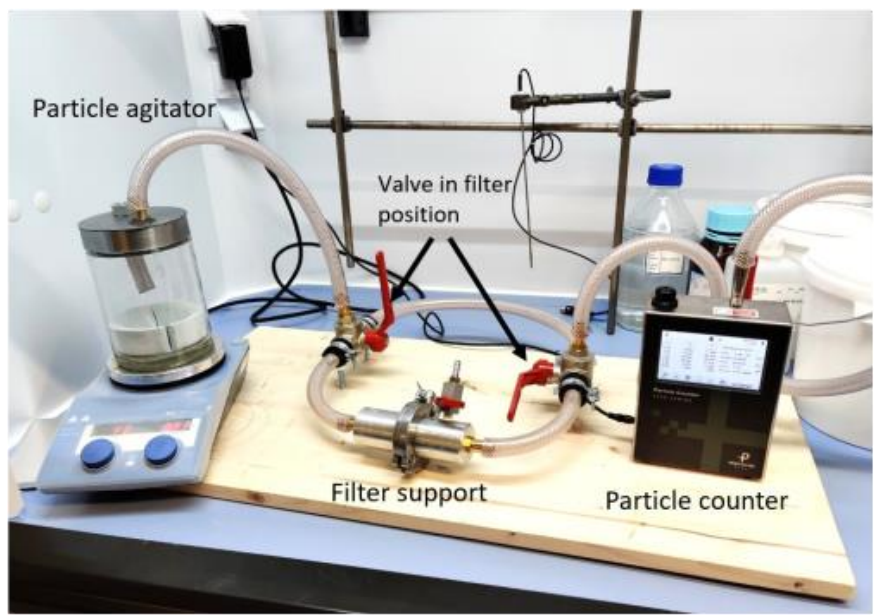

*Figure S4. Image of the custom-made filter characterization bench.*

### Nanoindentation results for polymer 7

Nanoindentation tests have been performed on selected polymer **7** filters with different deposition time and the two reference facemasks. These tests did not show any significant dependence of the mechanical stiffness on the electrospinning deposition time (not shown). Figure S5 shows the results of the elastic modulus of individual filter fibres for PEA-polymer **7** (20 min deposition time) and the two PP based commercial filters. The electrospun fibres exhibited an elastic modulus E=4.0 +/- 0.4 GPa which is by approximately a factor two greater than the commercial PP filters (E=1.84 +/- 0.2 GPa for Aspop and 2.3 +/- 0.2 GPa for Facemask). This nanomechanical test, in terms of mechanical properties, demonstrates that electrospun layers could partially compensate their lower layer thickness with respect to commercial filters through higher fibre stiffness.

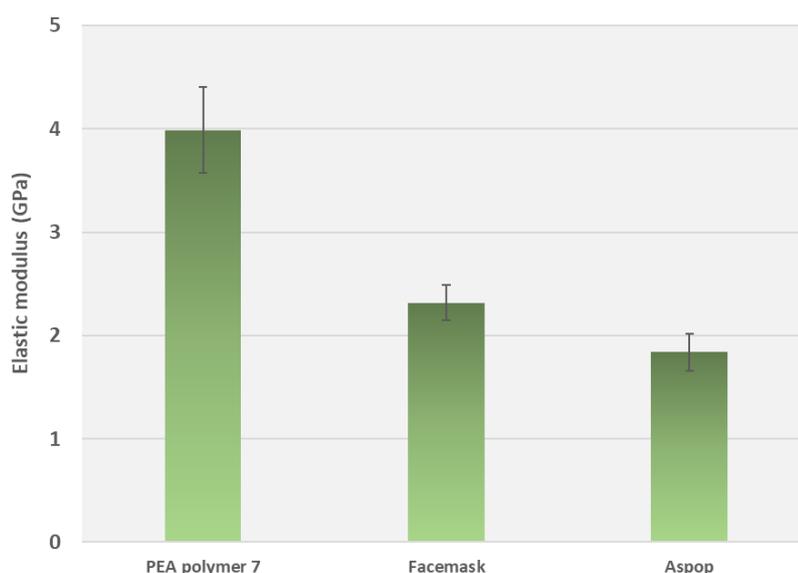

*Figure S5. Elastic modulus of electrospun polymer 7 with 20 min deposition time and commercial masks as determined by nanoindentation. The electrospun PEA has shown an elastic modulus that is approximately a factor of two greater than PP-based commercial fibres.*

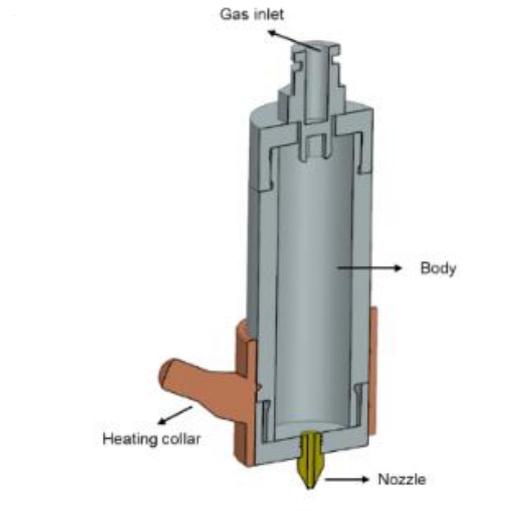

**Figure S6**. Extrusion head for the melt-spinning process